\documentstyle[twoside,fleqn,espcrc2,epsfig]{article}

\newcommand{\AmS}{{\protect\the\textfont2
  A\kern-.1667em\lower.5ex\hbox{M}\kern-.125emS}}

\title{Astroparticle Physics: \\
       The High Energy Tail of the Cosmic Ray Spectrum}

\author{Enrique Zas\address{Departamento de F\'\i sica de Part\'\i culas,\\
Universidade de Santiago de Compostela, E-15706 Santiago, Spain.\\
zas@fpaxp1.usc.es}\thanks{This work was supported in part by the European 
Science Foundation (Neutrino Astrophysics Network N. 86), by the CICYT (AEN99-0589-C02-02) and by Xunta de Galicia (PGIDT00PXI20615PR).}}

\begin{document}


\begin{abstract}

In this article I review the main theoretical problems that are posed by 
the highest energy end of the observed cosmic ray spectrum, stressing the 
importance of establishing their composition in order to decide between proposed scenarios. 
I then discuss the possibilities that are opened 
by the detection of inclined showers with extensive air shower arrays. 
Recent progress in modelling magnetic deviations for these showers has 
allowed the analysis of inclined showers that were detected by the Haverah 
Park experiment. This analysis disfavours models that predict a large 
proportion of photons in the highest energy cosmic rays and open up new 
possibilities for future shower array detectors particularly those, like 
the Pierre Auger Observatory, using water \v Cerenkov detectors. 

\end{abstract}

\maketitle

\section{HIGH ENERGY COSMIC RAYS}


%
%
%
%

Cosmic rays are elementary particles arriving at the Earth from outside 
that were discovered in the beginning of the 20th century as one of the 
main sources of natural radiation. 
The cosmic ray spectrum has been observed 
as a continuum at all energies since their discovery.
Throughout this period cosmic rays have always been the source 
of the highest energy elementary particles known to mankind, and 
for this reason they have given birth to particle physics. 
The high energy tail of the spectrum as it is known today 
corresponds to energies up to 3~$10^{20}~$eV and rates of a few 
particles per km$^2$ per century. 

It is remarkable that the cosmic rays have a quite featureless 
power law energy spectrum which decreases as approximately the cube 
of the primary energy. For energies above the few hundred TeV 
the observed flux necessarily requires techniques 
that take advantage of the extensive air showers that the arriving 
particles develop as successive secondary particles cascade down 
into the atmosphere. 
Shower measurements allow the reconstruction of the arrival directions 
and the shower energy but the nature of the primary particle 
is extracted by a number of indirect methods. 


For energies above few tens of GeV the detected particles, mainly protons, 
have arrival directions with a remarkably isotropic distribution. This is 
understood in terms of diffusive propagation in the galactic magnetic fields. 
As the energy rises above a given value that depends on the charge of 
the particle, propagation in the Galaxy should cease to be diffusive. 
Such high energy particles are expected to be extragalactic. 

The observation of high energy cosmic rays 
has been recently reviewed by Nagano and Watson \cite{WatsonNagano} 
who have shown that there is very good agreement between different 
experiments including the low and high energy regions of the spectrum. 
There is increasing evidence for a different component of the high energy 
end of the cosmic ray spectrum \cite{gaisser}. 
Combining data of five different experiments, 
AGASA, AKENO, Haverah Park, Stereo Fly's Eye and 
Yakutsk, Nagano and Watson conclude that there is a clear signal of a change of 
the spectral slope in the region just above $10^{18}~$eV \cite{WatsonNagano}. Composition studies have also 
given indications that there is a change to light element composition for energies above $\sim 10^{18}~$eV  \cite{composition} although this 
conclusion is model dependent to some extent  \cite{gaisser}. 
Also the small anisotropy ( $4 \%$) of $10^{18}~$eV cosmic rays in 
the direction of the galactic anticenter detected with AGASA disappears 
at higher energies \cite{hayashida}. 

The highest energy events detected present a serious challenge 
to theory and little is known about their origin. 
If they are protons they should attenuate in 
the Cosmic Microwave Background (CMB) over distances of order 50~Mpc. 
Such attenuation 
was predicted to appear in the cosmic ray spectrum as a cutoff, the 
Greisen-Zatsepin-Kuz'min (GZK) cutoff, just above 4~$10^{10}~$GeV \cite{GZK}. 
If they are photons or iron nuclei it 
turns out that interactions with the radio and the infrared backgrounds 
are respectively responsible for attenuations over similar or even shorter 
distances. 
No such features are seen in the observed cosmic ray spectrum. 
If they are produced sufficiently close to us to avoid the cutoff then 
the arriving particles should be pointing to their sources. This seems 
difficult to accommodate because there are very few known astrophysical 
sources capable of reaching the observed energies and on the other hand 
there is little evidence for the anisotropy that would result. 

This article firstly discuses the problem presented by the high energy 
end of the cosmic ray spectrum with emphasis in the role of composition. 
Then it outlines new progress made in understanding different 
features of inclined showers illustrating how these showers can contribute 
to the composition issue reviewing the results obtained by a recent 
analysis of the inclined data in Haverah Park. 


\section{ORIGIN: AN UNSETTLED ISSUE}

\begin{figure}[htb]
\centerline{\epsfig{file=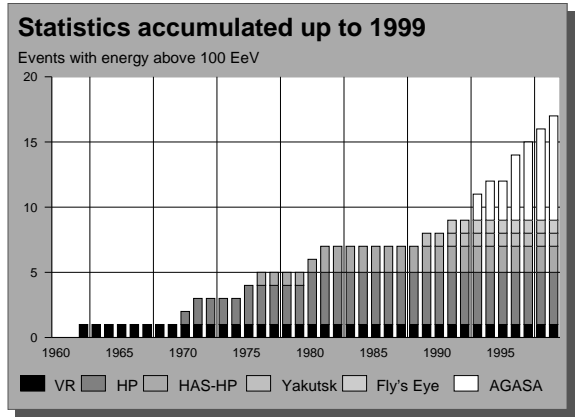,height=2.2in,width=3.in}}
\caption{Accumulated events of energy exceeding $10^{20}$ eV 
plotted as a function of time as detected by different experiments: 
Volcano Ranch (VR), Haverah Park (HP), Horizontal Air Showers in 
Haverah Park (HAS-HP), Yakutsk, Fly's Eye and AGASA.}
\label{uptonow}
\end{figure}

The discovery of events with energies above $10^{20}$~eV (100 EeV) 
dates back to the 1960's, to the early days of air shower detection 
experiments \cite{EeVents}. Since then they have been slowly but 
steadily detected by different experiments as illustrated in 
Fig.~\ref{uptonow}. Now there is little doubt about the non 
observation of a GZK cutoff, with over 17 published events above 
$10^{20}$~eV and five preliminary new events from HiRes \cite{hires}. 
On the contrary the data suggests that the spectrum continues 
smoothly within the statistical errors, possibly with a change 
of slope. On the other hand the data show no firm evidence of 
anisotropy but the significance of such studies is even more 
limited by the poor statistics. 

Both the details of the spectrum at the cutoff region and the extent 
to which the arrival directions of these particles cluster in the 
direction of their sources are very dependent on a number of 
unestablished issues. These include the 
source distribution, the distance of the nearest sources, 
their emission spectra, the intervening magnetic fields and of course 
on the nature of the the cosmic rays themselves or composition. 
If these particles are nuclei or photons the observational evidence 
is suggesting that these are coming from relatively nearby sources 
compared to the 50~Mpc scale. The conclusive power of observations is 
however strongly limited by both the poor statistics and a 
complex interrelation of hypotheses, but the situation is bound 
to change in the immediate future with a new generation of large 
aperture experiments, some like HiRes \cite{hires} already in operation, 
others in construction \cite{auger} and many others in 
planning \cite{EUSO,zaspuebla}.
The complex puzzle that connects particle physics, 
magnetic fields, and cosmic rays has attracted the attention of many fields 
in physics. 

\begin{figure}[htb] 
\centerline{\epsfig{file=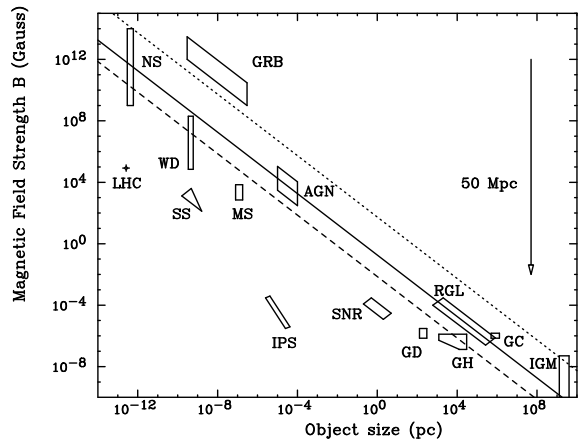,height=2.3in,width=3.in}}
\caption{Hillas Plot of the typical size of a possible accelerator $L$ versus 
its magnetic field strength $B$. From upper left to lower right the astrophysical objects correspond to Neutron Stars, White Dwarfs, Sun Spots, 
Magnetic Stars, Gamma Ray Bursts, Active Galactic Nuclei, 
Inter Planetary Space, Supernovae Remnants, Radiogalaxy Lobes, 
Galactic Disk, Galactic Halo, Clusters of Galaxies and the Inter Galactic 
Medium. Also shown is the point corresponding to the largest accelerator 
in planning LHC. The straight lines  
represent the limits given by Eq~(13) for protons (full), Iron nuclei (dashed) and for protons assuming a $10\%$ efficiency (dots).}
\label{hillasplot}
\end{figure}

In a conventional approach these particles would be nuclei as the 
bulk of the cosmic ray spectrum which are accelerated through 
stochastic acceleration as suggested by Fermi in 1949. 
This happens every time 
charged particles cross interfaces between regions that have astrophysical 
plasmas with different bulk motions, such as shock fronts. 
Transport is assumed to be diffusive in the plasma's magnetic field 
and on average in these processes a very small fraction of the bulk 
plasma kinetic energy is transferred as a boost to the individual 
particles, that typically end up with a power like spectrum. 
Acceleration of a particle of charge $Ze$ to an energy $E$ is strictly 
limited by dimensional 
arguments to objects that are sufficiently large or have sufficiently 
large magnetic fields. Basically for a particle with momentum $p$ to be 
able to undergo such a boost, propagation must be diffusive, or 
equivalently the accelerator region $L$ must be larger than the Larmor 
radius of the particle, $R$, in its characteristic magnetic field $B$: 
\begin{equation}
R={p \over Z e B} < L    \;\;\;\;\;  E\simeq pc < ZeBcL
\end{equation}
The requirement is well known by accelerator designers and is the 
ultimate reason for their high cost. 
It turns out that few of the 
known astrophysical objects satisfy the minimum requirements to accelerate 
particles to $10^{20}~$eV. This is conveniently illustrated in 
a plot first conceived by Michael Hillas \cite{Hillas} which is reproduced 
in Fig.~\ref{hillasplot}. 
A number of possible scenarios are being discussed; 
they imply acceleration in some objects including young pulsars, 
Gamma Ray Bursts (GRB), our own galaxy, active galaxies 
and the local group of galaxies \cite{review}. 
The power supply needed to keep the 
observed cosmic rays at the highest energies is consistent with the 
known power and distributions of these objects \cite{gaisser}. 

It is difficult to explain the observed flux spectrum 
in this conventional approach. 
A solution in which particles are accelerated nearby has difficulties 
because there are very few objects which are capable of 
accelerating particles to the maximum observed energies. 
Moreover many such objects are either too 
large or too distant for the cosmic ray spectrum detected at the Earth 
not to show the predicted GZK cutoff. If the sources were to be galactic 
no absorption cutoff would be expected but some spectral 
features are predicted for primary protons that are produced at 
a distance of more than a few Mpc. 
On the other hand 
the non observation of anisotropy complicates the puzzle, because the 
location of the possible accelerators in our vicinity is pretty well known.  
Primary protons having energies in the $10^{20}$~eV range are expected 
to be little deviated in the galactic magnetic fields. Our knowledge of 
extragalactic magnetic fields is poor but 
bounds on extragalactic magnetic fields 
also imply that the deviations of protons produced in the few Mpc 
range are not large. There are however possible configurations 
of the extragalactic magnetic fields that could explain many of the 
ultrahigh energy events as coming from a single source \cite{tanco}. 
The issue is far from being resolved and knowledge about composition 
is bound to play a crucial role for future progress in understanding. 

Motivated by particle physics beyond the standard 
model, many alternatives have been proposed that avoid acceleration and 
others that postulate different particles or different interactions. 
These include annihilation of topological 
defects created in the early universe, heavy relics that survive 
from the primeval bath, non thermal particles that couple to gravity, 
or wimpzillas and annihilation of relic neutrinos with messenger 
neutrinos coming from remote places \cite{review}. 
As regards composition two large categories of possible scenarios 
can be made namely those in which the observed particles are 
accelerated and those 
in which they are decay products of other particles. These 
two classes differ greatly in composition. 
A knowledge of composition is doubly important because firstly it may 
decide between these two classes of solutions and secondly because it 
would simplify the task of interpreting anisotropy measurements. 

\section{COMPOSITION}

The models that depend on acceleration can reach higher energies 
if the accelerated particles have large charge $Z$. This shows as a 
different restriction line in Fig \ref{hillasplot}. The relative 
composition of different nuclei resulting from such a scenario 
will depend on the local abundances of the different nuclei and 
on the energy. Depending on distance to the source and the surrounding 
environment there may be energy losses, absorption and the production 
of secondary particle fluxes. For instance in Active Galactic Nuclei 
(AGN) models the accelerated protons are expected to interact with ambient 
light or matter to produce pions that decay into photons and neutrinos. 
The neutrinos can reach the Earth unattenuated and provide a signature 
of proton acceleration. Unless the environment becomes opaque to 
protons the relative fluxes of neutrinos and protons that reach the 
Earth should be a number of order one or smaller, just because 
neutrinos are secondaries with repect to protons. The relative fluxes 
of photons and protons would be similar to neutrinos or smaller 
depending on the photon absorption both at the source and during 
transport to Earth. Ratios of the same order of magnitude would apply 
to most acceleration models.

Most of the non accelerating alternatives postulate the cosmic rays are 
products of the decay of other more massive particles produced by 
different mechanisms. Typically these particles of mass of 
$10^{24-25}~$eV (often an $X$ particle) decay into standard 
model particles which eventually fragment into hadrons, mostly pions 
and a small fraction of order $3\%$ of nucleons. 
While neutral pions decay into photons charged pions decay into neutrinos. 
Fragmentation processes, known form accelerator experiments and 
extrapolated to the high energies, become the common reference point 
for these mechanisms. 
For this reason all these models share a very similar composition 
dominated by photons and neutrinos which typically are about ten times 
more numerous than nucleons at the production site. 

Depending on the source distribution the relative fluxes of 
these particles are modified 
through their interactions with the background radiation fields. The 
neutrinos are the particles that preserve their production spectrum 
without being attenuated. Protons get attenuated in few tens of Mpc 
in the cosmic microwave background, (the GZK cutoff), while photons 
are attenuated already in few Mps mainly through pair production in the 
radio background. As a result the ratio of neutrinos to protons can 
in principle become higher at the Earth than when they are produced 
if the sources are quite distant or cosmologically distributed. 
Many of 
the proposed mechanisms are expected to cluster in our galactic halo. 
This possibility is receiving a lot of attention because it would provide 
a relatively natural explanation for the absence of the GZK cutoff. 
In that case however the sources will be quite near and the ratio of 
photons to nucleons should be expected to be of order 10, close to its 
value at production. Other sources are not expected to cluster and hence 
the photon to nucleon ratio is expected to drop to values close to one. 
The ratio of photons to nucleons depends on 
the source distribution and is rather sensitive to clustering. 


\section{INCLINED SHOWER FEATURES}

Most air shower detectors in existence consist on arrays of particle 
detectors that sample the extensive air shower front as it reaches the 
ground. Multiple particle production takes place in the successive 
high energy interactions produced as the shower penetrates the medium. 
As a result the number of particles in the shower front increases exponentially. 
When the average particle energy in the front becomes too low for multiple 
particle production the shower reaches it maximum number of particles. 
The development of these showers is typically governed by the 
radiation length in the material which is of order 36~g~cm$^{-2}$ in air 
and shower maximum, which is only logarithmically dependent on the primary 
particle energy, occurs at a couple of thousand meters for vertical showers 
of energies of order $10^{20}$~eV. 

Vertical showers are close to shower maximum when reaching the Earth's 
surface, have pretty good circular symmetry and are less 
affected by the Earth's magnetic field. 
It is thus not surprising that air showers have traditionally been studied at close to vertical incidence, typically for zenith angles below 
$45 ^\circ$, in summary because it is much simpler. 
Moreover in most extensive 
air shower arrays the particle detectors are oriented to have maximum 
collection area for vertical incidence. Since these detectors are often 
scintillator sheets, they tend to become very inefficient for very inclined 
showers. 

As the zenith angle increases the traversed atmospheric depth rises from 1000 
to close to 36000 g~cm$^{-2}$. As a result the shower maximum is reached in 
the upper layers of the atmosphere and most of the shower is absorbed before 
reaching the ground. It has been known for a long time that weakly interacting 
particles such as neutrinos can induce close to horizontal air showers deep 
in the atmosphere with particle distributions that are quite similar to 
vertical showers \cite{berezinskii,hsvum}. Air shower array detectors 
looking in the close to horizontal direction can thus be sensitive to high energy neutrino fluxes \cite{capelle}. 
In fact most bounds on neutrino fluxes have already been obtained from 
air shower experiments \cite{blancoPRL,baltrusaitis}. 

The original motivation of studying inclined showers was to understand the cosmic ray background to the neutrino induced showers. Although the electromagnetic part of 
the air shower induced by an inclined cosmic ray is indeed absorbed before 
reaching ground level, the shower front however also contains muons which are 
mainly produced by charge pion decay when the primary particle is a hadron. 
These muons do travel practically unattenuated all the slant atmospheric depth 
and produce density patterns on the ground that are much affected by the 
Earth's magnetic field. It has recently become quite clear that such inclined 
showers can be analysed. This not only nearly doubles the aperture of any 
air shower array but, when combined with vertical measurements, it has a remarkable potential for the study of primary composition \cite{avePRL}. 

Much development in this field has been possible by the modelling of the muon 
density patterns produced by inclined showers under the influence of the 
Earth's magnetic field \cite{rate}. 
The lateral distributions of muons in inclined showers can be understood 
in terms of a simple model \cite{HSmodel} in which the magnetic field 
is firstly neglected. The model stresses two important facts that have 
been extensively checked with simulations in the absence of a magnetic 
field \cite{HSmodel}: 
Most of the muons in an inclined shower are produced in a well 
defined region of shower development which is quite distant from the ground 
and the lateral deviation of a muon is inversely correlated with its energy. 

Indeed most of the fundamental properties of these inclined showers are 
governed by the distance and depth travelled by the muons. 
It is remarkable that the average slant distance travelled by the muons 
is of order 4 km for vertical showers, becomes 16 km at 60$^\circ$ 
and continues to rise as the zenith angle rises to reach 300 km for 
a completely horizontal shower. This distance plays a crucial role 
as a low energy smooth cutoff for the muon energy distribution. 
For inclined showers the muons must have much more energy at production 
to reach ground level without decaying than in the vertical case. 
Both the travel time and the muon energy loss become relevant. 


The model simply assumes that all muons are produced at a given altitude 
$d$ with a fixed transverse momentum $p_\perp$ 
that is uniquely responsible for the muon deviation from shower axis. 
In the transverse plane to the shower at ground level the muon deviation, 
$\bar r$, is inversely related to muon momentum $p$. 
The density pattern has full circular symmetry when there is no magnetic 
field. When the magnetic field effects are considered 
the muons deviate a further distance $\delta x$ in the perpendicular 
direction to the magnetic field projected onto the transverse plane 
$\vec B_\perp$, given by:
\begin{equation}
\delta x = \frac {e \vert B_\perp \vert d^2} {2p} = 
\frac{0.15 \vert B_\perp \vert d}{p_{\perp}} \; \bar r = \alpha \; \bar r,
\label{alpha}
\end{equation}
where in the last equation $B_{\perp}$ is to be expressed in Tesla, 
$d$ in m and $p_{\perp}$ in GeV. 
As the muon deviations are small compared to $d$ they can be 
added as vectors in the transverse plane and the muon density pattern is 
a relatively simple transform of the circularly symmetry pattern. 
The muon patterns in the transverse plane can be projected onto the ground 
plane to compare with data as well as standard simulation programs. 

Eq.~\ref{alpha} is telling us that all positive 
(negative) muons that in the absence of a magnetic field would 
fall in a circle of radius $\bar r$ around shower axis, are translated a distance $\delta x$ to the right (left) of the $\vec B_{\perp}$ direction.   
The dimensionless parameter $\alpha$ measures the relative effect of the 
translation. For small zenith angles $d$ is relatively small and 
$\alpha << 1$ so that the magnetic effects are also small, and results 
into slight elliptical shape of the isodensity curves. 

For high zeniths however $\alpha >1$ the magnetic translation exceeds 
the deviation the muons have due to their $p_{\perp}$. 
In this case {\sl shadow} regions with no muons 
are expected in the muon density profiles. For an approximate 
$p_{\perp} \sim 200$~MeV and $B_{\perp} = 40~\mu$T this happens when 
$d$ exceeds a distance of order 30~km, that is for zeniths above 
$\sim 70^{\circ}$.  
These shadow regions in the transverse plane are indeed an outstanding 
feature of the ground density profiles at high zeniths as seen in 
the simulations. 

The simple model can be actually generalized to account for muon energy 
distributions as a function of distance to shower axis, and improved using 
the correlation between the average muon energy and the distance to shower 
axis as obtained in dedicated simulations. When all this is done the 
obtained muon density patterns are shown to be accurately reflect those 
obtained with simulations and this proves to be a very useful tool for 
the study of inclined showers. 

For each zenith angle the primary particle energy sets the normalization 
of the particle densities. For proton primaries the total number of muons in the shower scales with the proton energy $E$ as:
\begin{equation}
N=N_{ref}~E^\beta
\label{Escaling}
\end{equation}
where $\beta$ is a constant. 
It is remarkable that the shape of the lateral distribution of the muons does 
not significantly change for showers of energy spanning over three orders 
of magnitude. The same happens for heavier nuclei with slightly different 
parameters. The results are slightly model dependent. 
Two alternative hadronic interaction models have been compared, the Quark 
Gluon String Model (QGSM) and SIBYLL to give also the same behaviour with 
also different parameters. Table~\ref{nmu.tab} illustrates these effects. 

\begin{table}
\begin{center}
\begin{tabular}{|lrcc|} \hline
Model & A & $\beta$ & $N_{\mu}$ ($10^{19}$ eV) \\\hline\hline
SIBYLL & 1 & 0.880 & 3.3 10$^{6}$ \\
       & 56 & 0.873 & 5.3 10$^{6}$ \\ \hline
QGSJET & 1 & 0.924 & 5.2 10$^{6}$ \\
       & 56 & 0.906 & 7.1 10$^{6}$ \\ \hline
\hline
\end{tabular}
\end{center}
\caption{Relationship between muon number and primary energy for 
proton and irons in two hadronic models (see equation~\ref{Escaling}).}
\label{nmu.tab}
\end{table}

As a final result the muon distributions can be represented by continuous 
functions which are analytically obtained once we know the main features 
of a shower in the absence of magnetic field. In practice this implies that 
only different zenith angles have to be simulated. Different azimuths are 
obtained by adequate transformations of the showers without magnetic deflections. 
The algorithm is fast and allows detector simulation and 
also event by event reconstruction of data obtained by air shower experiments. 

\section{PHOTON COMPOSITION}

This powerful technique has been used to analyse the inclined shower 
data obtained in the Haverah Park array. The Haverah Park detector was 
a 12~km$^2$ air shower array using 1.2~m deep water \v Cerenkov 
tanks that was running from 1974 until 1987 in Northern England 
which has been described elsewhere \cite{haverah}. It is possibly the most 
appropriate detector for this study because the water \v Cerenkov tanks 
have a uniquely large cross section to sample shower fronts of 
horizontal air showers. Moreover the \v Cerenkov technique gives larger 
signals for muons than for electrons simply because the muons have typically larger energies and travel through the whole detector. 

A careful study has been made of the energy deposition of signal in water 
\v Cerenkov tanks by horizontal muons using conventional simulation 
programs for this purpose \cite{Wtank}. A number of effects have to be 
considered to interpret the observed data. Inclined particles 
can produce light that falls directly into the phototubes without being 
reflected in the tank walls. Horizontal muons produce more signal through 
delta rays because on average they have higher energies than in vertical 
showers. There is a significant signal deposited by electromagnetic particles 
that arise mainly through muon decay. Finally the higher energy muons are 
more likely to deposit more energy in the tanks because of catastrophic 
energy losses. The event rate as a function of zenith angle 
has been simulated with careful treatment of all these effects 
using the muon distributions obtained as described in the previous section. 
The qualitative behaviour of the registered rate is well described in the 
simulation and the normalization is also shown to agree with 
data to better than $30\%$ using the measured cosmic ray spectrum for 
vertical incidence, assuming proton primaries and using the QGSM 
model \cite{rate}. 

More impressive are the results of fits of the models for muon densities 
to the observed particle densities sampled by the different detectors on 
an event by event basis. The nearly 10,000 events recorded with zenith angles 
above $60^{\circ}$ have been analysed for arrival directions, impact point 
and primary energy in the assumption the primaries are protons. A complex 
sequence of arrival direction and density fits is performed to minimize 
the effect of correlations between energy and arrival directions. 

\begin{figure}
\centerline{\epsfig{file=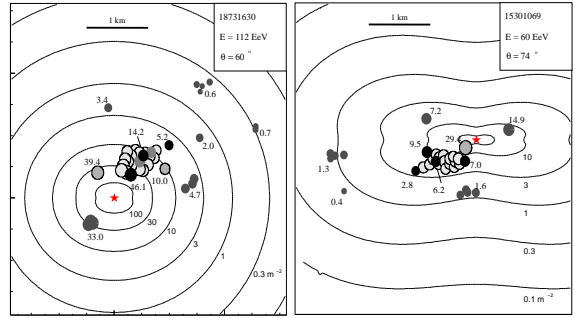,height=1.7in,width=3.in}}
\caption{Density maps of two events in the plane perpendicular 
to the shower axis. Recorded muon densities are shown as circles 
with radius proportional to the logarithm of the density. The detector
areas are indicated by shading; the area increases from white to black as
1, 2.3, 9, 13, 34 m$^{2}$. The position of the best-fit core is
indicated by a star. Selected densities are also marked. The y-axis 
is aligned with the component of the magnetic field perpendicular to the
shower axis.}
\label{events.fig}
\end{figure}
The analysed date is subject to a set of quality cuts: the shower 
is contained in the detector (distance to core less than 2~km), 
the $\chi ^2$ probability of the event is greater 
than $1 \%$ and the downward error in the reconstructed energy is less than 
$50 \%$. These cuts ensure that the events are correctly reconstructed and 
exclude all events detected above $80^\circ$. Examples of reconstructed 
events compared to predictions are illustrated in Fig.~\ref{events.fig}. 
Two new events with energy exceeding $10^{20}$~eV have been revealed. The results have been 
compared to a simulation that reproduces the same fitting 
procedure and cuts using the cosmic ray spectrum deduced from vertical air shower measurements in reference \cite{WatsonNagano}.
The agreement between the integral rate above $10^{19}$~eV 
measured and that obtained with simulation is striking when 
the QGSJET model is used. Sibyll leads to a slight underestimate 
\cite{avePRL}. 

The universality of the muon lateral distribution function 
is very powerful and once the equivalent proton energy is determined 
for all events, the corresponding energies in the assumption 
that the primaries are iron nuclei (photons) can be obtained 
multiplying the proton energy 
by a factor which is $\sim 0.7$ (6) for $10^{19}$~eV. 
As a result when a photon primary spectrum is 
assumed the simulated rate seriously underestimates the observed data 
by a factor between 10 and 20. A fairly robust bound on the 
photon composition at ultra high energies can be established assuming a
two component proton photon scenario. The photon 
component of the integral spectrum above $10^{19}~$eV (4~$10^{19}$~eV) 
must be less than $41\%$ ($65\%$) at the $95\%$ confidence level. 
Details of the analysis are presented in \cite{avePRL}. 

The results of this method when applied to a first analysis of inclined 
showers produced by cosmic rays above $10^{19}~$eV demonstrates that the 
study of inclined showers not only can double the acceptance of air shower 
arrays but it can be a very useful tool for the study of photon composition.

\end{document}